\documentstyle[aps,epsf,prl]{revtex} 

\begin{document}
\twocolumn
\title{Acceleration of quasi-particle modes in Bose-Einstein condensates}
\author{Karl-Peter Marzlin and Weiping Zhang}
\address{
School of Mathematics, Physics,
Computing and Electronics, Macquarie University, Sydney, NSW 2109,
Australia}
\maketitle
\begin{abstract}
We analytically examine the dynamics of quasi-particle modes
occuring in a Bose-Einstein condensate which is subject to a
weak acceleration. 
It is shown that the momentum of a quasi-particle mode is squeezed
rather than accelerated.
\end{abstract}

The achievement of Bose-Einstein condensation in magnetic traps
\cite{experimente} has led to a great interest in the
physical properties of inhomogeneous atomic quantum gases. 
In the current research, one of the central topics is to study
collective excitations of quasi-particle modes in
Bose-Einstein condensates (BEC). The frequency spectrum of elementary
excitations in a trapped $^{87}$Rb BEC has been experimentally measured
at JILA \cite{excexperim}. The propagation of sound waves in an
excited BEC has been demonstrated at MIT \cite{soundwaves}.
Theoretical studies have revealed the characteristics of
low-energy quasi-particle modes in such nonlinear systems
\cite{theorie}.

In this paper we are interested in another aspect of collective excitations
of BECs. We will study the dynamics of quasi-particle modes and
show theoretically how they are manipulated by a weak homogeneous
force. Since the dispersion relation of quasi-particle modes
is much different from that of ordinary Schr\"odinger particles
their dynamics will show up unusual effects like
an acceleration-induced squeezing of position or momentum. 
A squeezing of the position may correspond to a localization
of sound waves in quantum gases.

To initiate our idea, we start with a simplified model where a BEC
trapped in a box of size $L$. In practice the box may be realized
by a sufficiently anharmonic trap or an atomic cavity. We then add to
this trapping potential a second, linear potential which describes
the external homogeneous force. As a result, the complete potential
becomes asymmetric and should incorporate an area where the potential is
almost linear (see Fig.~1).
A harmonic trap is not suited for this purpose since the homogeneous
force then only leads to an overall displacement of the trap without
changing its shape.
 
The time evolution of
a BEC can be appropriately described by the 
time-dependent Gross-Pitaevskii equation,
\begin{equation} 
   i \hbar \frac{d \psi}{dt} = \left \{ \frac{p^2}{2M} +
   V(x) \right \}\psi +\kappa |\psi|^2 \psi \; ,
\label{nlse} \end{equation} 
where $\psi(x)$ is the normalized collective wavefunction of the BEC, 
$M$ the atomic mass, and $\kappa := 4 \pi \hbar^2 N 
a_{\mbox{{\scriptsize scat}}} /M$. 
$N$ is the number of condensed atoms and $a_{\mbox{{\scriptsize scat}}} >0$
the scattering length of the interaction between the atoms.
$V(x)$ is an external potential.

The ground state $\psi_g(x,t) = \exp [-i \mu t/\hbar ] \psi_g(x)$
of the trapped BEC  is a stationary solution of Eq.~(\ref{nlse}),
where $\mu$ denotes the chemical potential. 
$\psi_g(x)$ can be assumed to be real. For current experiments
the nonlinear interaction between the atoms is relatively strong
compared to the BEC's kinetic energy. It is then possible to
neglect the kinetic term in Eq.~(\ref{nlse}) and to describe the
ground state in the Thomas-Fermi approximation (TFA) \cite{tfapp},
\begin{equation} 
  \kappa |\psi_g|^2(x) = \mu - V(x) \; .
\label{tfa} \end{equation} 
Of course this approximation is only valid in areas where the
r.h.s.~(right-hand-side) does not become small or negative.

Elementary excitations around the ground state $\psi_g$ of Eq.~(\ref{nlse})
are described by writing the collective wavefunction
as $\psi(x,t) = \exp [-i \mu t/\hbar] \{ \psi_g(x) + \Theta (x,t) \}$ and
expanding Eq.~(\ref{nlse}) up to terms linear in $\Theta$. The resulting
equation,
\begin{equation} 
  i \hbar \frac{d \Theta}{dt} = R \Theta + S \Theta^* \; ,
\label{lindgl}\end{equation} 
with $R:= p^2/(2M) +V -\mu + 2 \kappa |\psi_g(x)|^2$
and $S:= \kappa \psi_g^2$, has for time-independent $R$ and $S$
the formal solution
\begin{eqnarray} 
  \Theta(x,t) &=& \left \{ \cos [ \hat{\omega}_+ t] -i (R+S)\frac{\sin
    [\hat{\omega}_+ t]}{\hat{\omega}_+} \right \} \mbox{Re}\Theta_0 +
    \nonumber \\ & &
    i  \left \{ \cos [ \hat{\omega}_- t] -i (R-S)\frac{\sin
    [\hat{\omega}_- t]}{\hat{\omega}_-} \right\} \mbox{Im}\Theta_0
\label{solution} \end{eqnarray} 
($\Theta_0 := \Theta(x,0)$). The formal frequency operators
\begin{equation} 
 \hbar^2 \hat{\omega}_\pm := \sqrt{(R\mp S)(R \pm S)}
\end{equation} 
correspond to the spectrum elementary excitations which is usually
calculated by using the Bogoliubov-DeGennes equations (see, e.g.,
Ref.~\cite{fetter72}). We remark that one can rewrite Eq.~(\ref{solution})
using $\hat{\omega}_+$ or $\hat{\omega}_-$ alone by exploiting relations
like $(R-S)\cos[\hat{\omega}_-t] = \cos [\hat{\omega}_+ t](R-S)$.

In areas where the TFA is valid the operators $\hat{\omega}_\pm$ can be
simplified to
\begin{equation} 
  \hbar \hat{\omega}_+ := \sqrt{\frac{\hat{p}^2}{2M}
  \left ( \frac{\hat{p}^2}{2M}
  +2 \mu - 2 V(\hat{x})\right )} \; ,
\end{equation} 
and a similar expression for $\hat{\omega}_-$.
If we restrict our considerations to quasi-particle modes with
small momentum we can neglect the second kinetic term in this
equation and arrive at
\begin{equation} 
  \hbar \hat{\omega}_+ := \sqrt{\hat{p}^2 \left (c^2 - V(\hat{x})/M\right )} 
  \; ,
\label{omplus} \end{equation} 
where we have defined the velocity of sound $c:= \sqrt{\mu/M}$. 
Obviously this approximation corresponds to the restriction to the linear
part of the dispersion relation, $\hbar \hat{\omega}_+ = c |\hat{p}|$,
for elementary excitations in a homogeneous, free BEC ($V(x)=0$).

Though Eq.~(\ref{omplus}) looks simple the square root on the 
r.h.s.~is operator-valued and in general difficult to analyze.
We therefore restrict our analysis to one spatial dimension and
consider an accelerating potential $V(x) = - M a x$ with
$a L \ll c^2$, where $a$ is the acceleration and $L$ the length of the
condensate. We then can expand $\hat{\omega}_+$ in $a$, leading to
\begin{eqnarray} 
  \hbar \hat{\omega}_+ &=& |\hat{p}| \left (1 + \frac{a \hat{x}}{2c^2} 
  - \frac{a^2 \hat{x}^2}{8 c^4} \right ) -i \frac{\hbar a}{4 c^2} 
  \mbox{sgn}(\hat{p}) 
  \nonumber \\ & &
  + i \frac{\hbar a^2}{8 c^4} \mbox{sgn}(\hat{p}) \hat{x} 
  + \frac{3 \hbar^2 a^2}{32 c^4} |\hat{p}|^{-1}+ O(a^3) 
  \nonumber \\
  \hbar \hat{\omega}_- &=& |\hat{p}| \left (1 + \frac{a \hat{x}}{2c^2} 
  - \frac{a^2 \hat{x}^2}{8 c^4} \right ) +i \frac{3\hbar a}{4 c^2} 
  \mbox{sgn}(\hat{p}) 
  \nonumber \\ & &
  - i\frac{3\hbar a^2}{8 c^4} \mbox{sgn}(\hat{p}) \hat{x} 
  + \frac{3 \hbar^2 a^2}{32 c^4} |\hat{p}|^{-1} + O(a^3)
\label{omexpand} \end{eqnarray} 
In the derivation we have used the commutation relation 
$[\hat{x},\mbox{ sgn}(\hat{p})] = 2 i \hbar {\cal P}_0$, where ${\cal P}_0$
is the projection operator on the state $p=0$.

Some caution is needed when expanding the square root of an operator as
in Eq.~(\ref{omexpand}). Already Eq.~(\ref{omplus}) is only meaningful
if $c^2 - V/M = c^2+ a x$ is larger than zero, in agreement with
the assumptions of the TFA. Hence our approach is only useful for
wavepackets that are sufficiently localized inside the interval
$ x \in [-c^2/a , -c^2/a + L]$. In addition we have to demand that
the quasi-particle momentum is not too small so that its 
De-Broglie-wavelength is not larger than $L$. This becomes evident in
the expansion (\ref{omexpand}) which diverges for $p \rightarrow 0$.
Mathematically the inverse of an operator with eigenvalue zero cannot
be defined. However, the inverse makes sense on a subspace that excludes
this eigenvalue, and our restriction to sufficiently large $p$
ensures that we are working in this subspace.
Our approach therefore allows us to study wavepackets which are localized
in phase space. For simplicity we will assume that the wavepacket
is localized around some momentum $p_0 > 0$ in momentum space so that
we can set the operator sgn($\hat{p}$) equal to one. This causes no loss of
generality as any wavepacket subject to the constraints given above
can be linearly split into a part with $p>0$ and one with $p<0$. The latter
case can be treated analogously to the case $p>0$ which we will consider
now.

Having derived an explicit expression for the frequency operators
we can return to Eq.~(\ref{solution}) and consider the
evolution of a localized elementary excitation. Obviously the time dependence
of $\Theta$ is essentially determined by the four exponentials
$\exp[\pm i \hat{\omega}_\pm t]$. Since $\hat{\omega}_-$ is very similar
to $\hat{\omega}_+$ we will only study the exponentials of $\hat{\omega}_+$
 which can be decomposed as
\begin{eqnarray}
  e^{-i \hat{\omega}_+ t} &=& 
  \exp \left \{  -\frac{at}{4c} + \frac{a^2 t^2}{16 c^2}  \right \} 
  \exp \left \{-i \frac{c t}{\hbar} \left ( 1+ \frac{a t}{4 c}
    \right ) \hat{p} \right \} 
  \nonumber \\ & & \times 
  \exp \left \{ -i \frac{a t}{2c\hbar} \left ( 1-\frac{at}{4c} \right )
     \hat{p} \hat{x}\right \}
  \exp \left \{ i \frac{a^2 t}{8 \hbar c^3} \hat{p} \hat{x}^2 \right \}
  \nonumber \\ & &
  \times \exp \left \{ \frac{a^2 t}{8c^3}\hat{x} \right \}
  \exp \left \{ -i\frac{3 \hbar a^2 t}{32 c^3} \hat{p}^{-1} \right \}
\label{expo} \end{eqnarray}
where the exponentials are correct to order $a^2$. The proof of
this relation is analogous to that of the Baker-Campbell-Hausdorf
theorem. For instance, an expression like $F(t) :=
\exp [ it( \lambda \hat{p} + \lambda^\prime \hat{p} \hat{x}^2) ]$ 
can be decomposed by making the ansatz 
$ F(t) = \exp [ i u(t) \hat{p} ] \exp [ i v(t) \hat{p} \hat{x}] 
\exp [i w(t) \hat{p} \hat{x}^2]$ and
differentiating both expressions for $F(t)$ with respect to $t$. Using
simple commutation relations the resulting equation corresponds to
a set of ordinary differential equations for $u(t)$, $v(t)$, and
$w(t)$ which can be solved exactly or, as is sufficient for our purposes,
to order $a^2$.

Eq.~(\ref{expo}) allows us to study in detail the time evolution of
a quasi-particle mode $\Theta(x,t)$ by discussing the action of
$\exp [\pm i \hat{\omega}_+ t]$ on $|\Theta_0 \rangle$
\cite{remark}, with
$\langle x |\Theta_0 \rangle  := \Theta_0(x)$. Before doing so we
remark that for $a=0$ we have $\hat{\omega}_-=\hat{\omega}_+$ and
$\exp [\pm i \hat{\omega}_+ t] = \exp [\pm i \hat{p} ct/\hbar ]$
(in accordance with the discussion given above
$p>0$ has been assumed).
The r.h.s. of this expression corresponds to a shift of the
position, $x^\prime = x \pm ct$. Hence, for $a=0$ the exponentials
$\exp [\pm i \hat{\omega}_+ t]$ describe the propagation of
sound waves to the right or to the left, respectively.
As the right-moving part can be obtained from the left-moving part
by time-inversion we will discuss the -left moving part
$\Theta_L(x,t) := \langle x | \exp[-i \hat{\omega}_+ t]| \Theta_0 \rangle$
only.
\vspace{1cm}
\begin{figure}[h]
\epsfysize=5cm
\hspace{2mm}
\epsffile{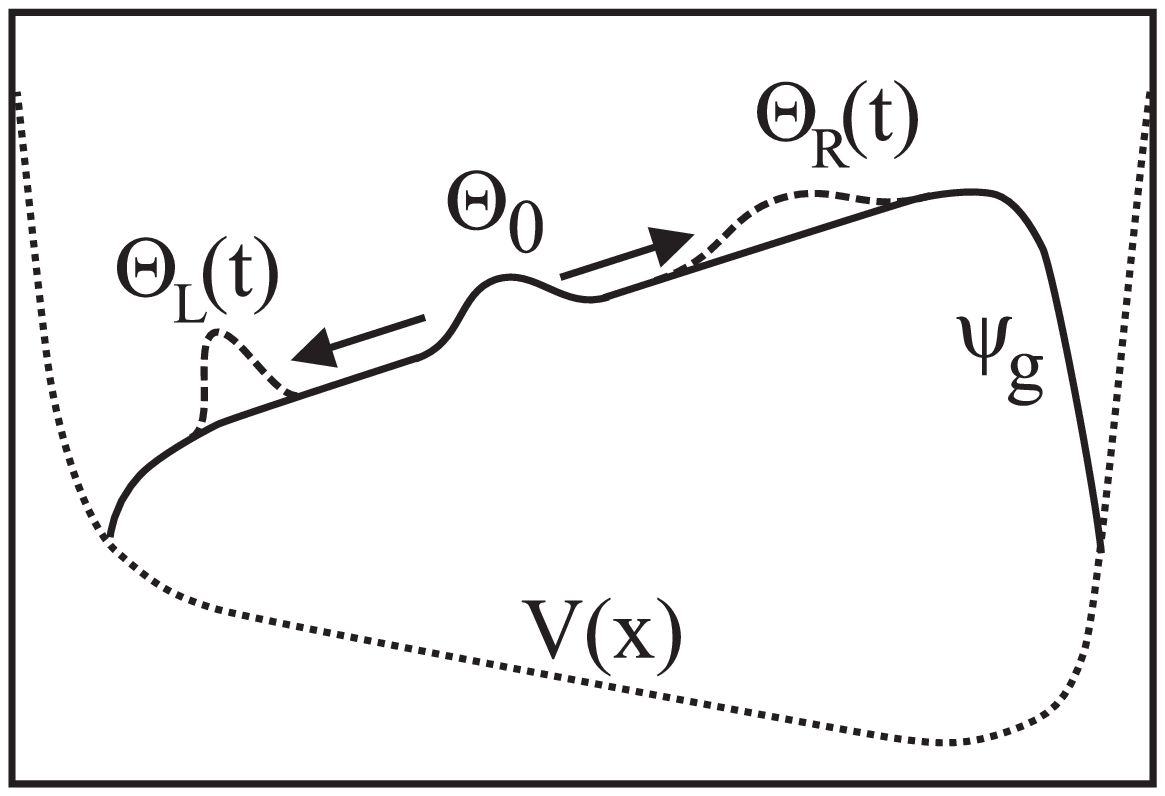}\vspace{-6mm}
\caption{Sketch of the evolution of a localized quasi-particle mode in the
linear regime of the combined potential $V(x)$. A small excitation $\Theta_0$
of the ground state $\psi_g$ is splitted into two counterpropagating
sound waves. The homogeneous force induces a squeezing operation on these
two wavepackets. The additional potential provided by $\psi_g$
leads to a reduction of the acceleration induced by the force.}
\end{figure}
To get an explicit expression for $\Theta_L(x,t)$ we have to analyze
the action of each individual exponential of Eq.~(\ref{expo}).
The first one is just a c-number and corresponds to a rescaling of
the wavefunction (note that Eq.~(\ref{lindgl}) does not conserve
the norm of $\Theta$). As discussed above the second exponential
corresponds to a time-dependent shift of the wavefunction and
describes an accelerated sound propagation:
\begin{equation} 
  \langle x  |  \exp \left \{ -i \frac{c t}{\hbar} 
  \left ( 1+ \frac{a t}{4 c} \right ) \hat{p} \right \} 
  | \phi \rangle 
  =  \langle x - ct - {\textstyle {1\over 4}} a t^2 | \phi \rangle \; .
\label{shift} \end{equation} 
In this equation $|\phi\rangle$ denotes an arbitrary element of
the Hilbert-space L$^2$({\bf R}) of square integrable functions.   
The third exponential is of the form $\exp [-i \alpha \hat{p} \hat{x}]$ with 
$\alpha := at[1-at/(4c)]/(2\hbar c)$. 
This operator corresponds to a squeezing transformation,
\begin{equation} 
  \langle x | e^{-i \alpha \hat{p} \hat{x}} | \phi \rangle = e^{-\hbar \alpha}
  \langle x e^{-\hbar \alpha} | \phi \rangle \; .
\label{squeez} \end{equation} 
A quick way to see this is to write the position and momentum operator
in terms of creation and annihilation operators as for the harmonic oscillator:
$\hat{x}\sim \hat{a}+\hat{a}^\dagger $, $\hat{p} \sim i (\hat{a}^\dagger 
-\hat{a})$. The exponential then can be
written as $\exp \{ \alpha [(\hat{a}^\dagger)^2 - \hat{a}^2 -1 ] \}$ 
which is just the
usual form of a single-mode squeezing transformation (see, e.g.,
Ref.~\cite{wallsbook}). Mathematically, Eq.~(\ref{squeez}) can be proven
by differentiating $f(\alpha,x) := \langle x | e^{-i \alpha \hat{p} \hat{x}} |
\phi \rangle$ with respect to $\alpha $.
the resulting differential equation, $\partial f/\partial \alpha =
-\hbar \partial (x f)/\partial x$ has, for the inital value
$f(0,x)= \langle x | \phi \rangle  $, the solution (\ref{squeez}).
Analogously one can show that in momentum space the squeezing operator
acts like
\begin{equation} 
  \langle p | e^{-i \alpha \hat{p} \hat{x}} | \phi \rangle =
  \langle p e^{\hbar \alpha} | \phi \rangle \; .
\label{squeez2} \end{equation} 
As usual the exponent of the squeezing factor
$\exp [\hbar \alpha ]$ has the opposite sign compared to the squeezing
of the conjugate variable $\hat{x}$. New compared to the usual squeezing of
ordinary particles is the c-number factor of 
$\exp [ -\hbar \alpha ] $ appearing
on the r.h.s. of Eq.~(\ref{squeez}). This factor and the resulting
asymmetry between Eqs.~(\ref{squeez}) and (\ref{squeez2})
is a consequence of the operator ordering: $\hat{p} \hat{x}$ instead of
$(\hat{p}\hat{x} + \hat{x}\hat{p})/2$. The order of $\hat{p} \hat{x} $ itself 
is a consequence of the asymmetric definition
of $\hat{\omega}_+=[\hat{p}^2(c^2+a \hat{x})]^{1/2}$. 

In an analogous treatment  one can show that the fourth exponential, 
which is of the form $\exp [ i \beta \hat{p} \hat{x}^2 ]$
with $\beta := a^2 t/(8 \hbar c^3)$, corresponds 
to a rational transformation of $x$,
\begin{equation} 
  \langle x | e^{i \beta \hat{p} \hat{x}^2} | \phi \rangle  =
  \frac{1}{(1-\hbar \beta )^2} \left \langle \left .
  \frac{x}{1-\hbar \beta x} \right |\phi \right \rangle \; .
\label{rational} \end{equation} 
The singularity at $ x= 1/(\hbar \beta)$ is unphysical since it
only appears if the wavefunction is not localized inside the area where
$c^2 \gg a x$ holds. The corresponding equation in momentum space is
much more complicated and difficult to interpret. We can gain some insight
into this case by considering the Heisenberg-picture evolution of the
momentum operator:
\begin{equation} 
  e^{-i \beta \hat{p} \hat{x}^2} \hat{p} e^{i \beta \hat{p} \hat{x}^2} = 
  \hat{p} (1+ \hbar \beta \hat{x})^2 \; .
\label{rational2}\end{equation} 
Since the momentum operator gets multiplied with the conjugate operator 
$\hat{x}$ its action in momentum space is difficult to interpret.
However, it still remains an operator linear in $\hat{p}$ so that its action
in position space is easy to understand: the wavefunction gets
multiplied with $(1+\hbar \beta x)^2$ and is then differentiated. 

The fifth exponential in Eq.~(\ref{expo}) acts as a position-dependent
rescaling of the wavefunction. In momentum space it corresponds to a
{\em  imaginary} shift of the momentum. The last exponential,
which is of the form $\exp [-i \gamma \hat{p}^{-1}]$
with $\gamma := 3 \hbar a^2 t/(32 c^3)$,  does not
change the momentum of the wavefunction but affects the quasi-particle's
position according to $\exp [-i \gamma \hat{p}^{-1}] \hat{x} 
\exp [i \gamma \hat{p}^{-1}] = \hat{x}+\gamma \hbar \hat{p}^{-2}$. 
This expression is not very insightful, however,
so that we exploit our assumption that the quasi-particle mode
is initially localized around some momentum $p_0$. We then can expand
the operator $\hat{p}^{-1}$ as $1/p_0 - (\hat{p}-p_0)/p_0^2 + 
(\hat{p}-p_0)^2/p_0^3 + \cdots$
so that we essentially get expressions which are linear or quadratic in
$\hat{p}$. The linear part shifts the position analogous to Eq.~(\ref{shift}). 
The term quadratic in $\hat{p}$ has the same structure as the
operator $\hat{p}^2/(2M)$ of the kinetic energy of an ordinary
Schr\"odinger particle. Hence an acceleration introduces a
momentum-dependent mass $\propto p_0^3/a^2$
for sound waves in a BEC.

Putting everything together by applying successively all
exponentials of Eq.~(\ref{expo}) to the quasi-particle mode
we arrive at
\begin{eqnarray} 
  \langle x | e^{-i \hat{\omega}_+ t} | \Theta_0 \rangle  &=&
  \left \{ 1 - \frac{3 a t}{4 c} + \frac{3 a^2 t x}{8 c^3} +
  \frac{3 a^2 t^2}{32 c^2} \right \}
  \nonumber \\ & & \times
  \exp \left [ -i 
  \frac{3 a^2 t\hbar}{16 c^3 p_0} \right ] \langle x_H(t) | \Theta_0 \rangle
\label{evolut} \end{eqnarray} 
with
\begin{eqnarray} 
  x_H(-t) &:=& \left ( 1- \frac{a t}{2 c}\right ) x + {1\over 4} a t^2
  -ct \left (1- \frac{3 a^2 \hbar^2}{32 c^4 p_0^2}\right )
  \nonumber \\ & &
  + \frac{a^2 t}{8 c^3} x^2 + O(a^3) + O(1/p_0^3) \; .
\label{xh} \end{eqnarray} 
This expression coincides with the Heisenberg-picture position operator
$\hat{x}_H(-t) =
\exp[-i \hat{\omega}_+ t] \hat{x} \exp [ i \hat{\omega}_+ t]$ if the
variable $x$ on the r.h.s. of Eq.~(\ref{xh}) is replaced by the 
position operator $\hat{x}$.

In view of the discussions given above the result (\ref{evolut})
is easy to understand. The prefactor in curly brackets corresponds to
a position dependent rescaling of the wavefunction, a combined
effect of the first exponential in Eq.~(\ref{expo}) and the prefactors
present in Eqs.~(\ref{squeez}) and (\ref{rational}). The phase factor
stems from the last exponential.
In $x_H(-t)$ the first term is a result of the squeezing transformation
(\ref{squeez}). The second term describes an acceleration of the
quasi-particle motion. Since the quasi-particle has to overcome
the potential created by the accelerated ground-state, which according
to Eq.~(\ref{tfa}) has the opposite sign as $V$ itself, the
acceleration is only half as large as for a usual particle
(prefactor 1/4 instead of 1/2). The term proportional to $c t$
corresponds to the sound propagation, where a correction
$\propto a^2/p_0^2$ to the velocity of sound appears because
of the last exponential in Eq.~(\ref{expo}). The last term, which
is nonlinear in $x$, is caused by the rational transformation of
Eq.~(\ref{rational}).

It is also interesting to study the
momentum operator $p_H(-t) :=
\exp[-i \hat{\omega}_+ t] \hat{p} \exp [i\hat{\omega}_+ t]$ 
in the Heisenberg picture, which is given by
\begin{equation} 
  p_H(-t) = \hat{p}\left ( 1+\frac{a t}{2c} + \frac{a^2 t^2}{4c^2} \right )
  - \frac{a^2 t}{4 c^3} \hat{p} \hat{x} + i\hbar \frac{a^2 t}{8 c^3} +O(a^3)
\label{ph} \end{equation} 
The first term describes, for $a<0$, a squeezing of the momentum. 
The second term
has its origin in the rational transformation (\ref{rational2}) which
also produces a higher order correction to the first term. The last
term is produced by the fifth exponential in Eq.~(\ref{expo}). Because of
this term $p_H(-t)$ is not a hermitean operator anymore. This is also a
consequence of the asymmetric occurence of $\hat{x}$ and $\hat{p}$ in 
the definition of $\hat{\omega}_+$.

For an ordinary accelerated particle the momentum in the Heisenberg
picture is given by $\hat{p}+Mat$ where the last term corresponds to
the increase of the momentum due to the acceleration. Notably this
term is absent in the case (\ref{ph}) of a quasi-particle. Thus,
the momentum of a quasi-particle is squeezed rather than accelerated by
a weak external homogeneous force. 

This behaviour can be understood if we consider $-\hat{\omega}_+$ 
of Eq.~(\ref{omplus}) as a kind of Hamiltonian for a classical
system, i.e., $H= -p [c^2+ax]^{1/2}$
(the minus sign arises because we are dealing with $x_H(-t)$ instead of
$x_H(t)$). 
It then becomes obvious that the classical canonical momentum
is not proportional to $dx/dt$ anymore so that its time evolution
is more complicated. This is reflected by the squeezing transformation.
We also remark that the results (\ref{xh}) and (\ref{ph}) also
serve as a further justification for Eq.~(\ref{omexpand})
since for $\hbar =0$ they do agree with the solutions of the
classical equations of motion, $\dot{x} = \partial H /\partial p$
and $\dot{p} = - \partial H/\partial x$.

Up to now we have analysed the time evolution of the left-moving
wavepacket
$\Theta_L(x,t) := \langle x | \exp[-i \hat{\omega}_+ t]| \Theta_0 \rangle$.
The corresponding right-moving part
$\Theta_R(x,t) := \langle x | \exp[i \hat{\omega}_+ t]| \Theta_0 \rangle$
can be obtained by reversing the sign of $t$ in the results
(\ref{evolut}), (\ref{xh}), and (\ref{ph}). It then becomes obvious that
(for $a>0$) the momentum of $\Theta_R$ is squeezed instead of its position
(see Fig.~1).

In conlusion we have demonstrated that a weak homogeneous force
acting on a BEC alters the motion of quasi-particle modes in a
different way than for ordinary particles. The acceleration of
the position is smaller since the mode has to overcome the
additional potential provided by the ground state. The momentum
is squeezed rather than accelerated because the ``Hamiltonian''
$\hat{\omega}_+$ contains a term $\propto \hat{p}\hat{x}$ 
which acts in the same way as a squeezing operator in quantum optics.

{\bf Acknowledgement}: The work has been supported by the
Australian Research Council.

\end{document}